\documentclass[
    ,aps
    ,groupaddress 
    ,reprint
    ,prstper
    ,citeautoscript
  ]{revtex4-1}

\usepackage{graphicx}
\usepackage{dcolumn}
\usepackage{bm}
\usepackage{color}
\usepackage{amsmath}
\usepackage{units}
\usepackage{needspace}
\usepackage{rotating}

\begin{document}

\title{Experts' understanding of partial derivatives\\ using the Partial
  Derivative Machine}

\author{David Roundy}
\affiliation{Oregon State University, Corvallis}
\email{roundyd@physics.oregonstate.edu}

\author{Allison Dorko}
\affiliation{Oregon State University, Corvallis}
\email{dorkoa@onid.oregonstate.edu}

\author{Tevian Dray}
\affiliation{Oregon State University, Corvallis}
\email{tevian@math.oregonstate.edu}

\author{Corinne A. Manogue}
\affiliation{Oregon State University, Corvallis}
\email{corinne@physics.oregonstate.edu}

\author{Eric Weber}
\affiliation{Oregon State University, Corvallis}
\email{Eric.Weber@oregonstate.edu}

\newcommand\myderiv[3]
  {\ensuremath{\left(\dfrac{\partial #1}{\partial #2}\right)_{#3}}}
\newcommand\myderivinline[3]
  {\ensuremath{(\slfrac{\partial #1}{\partial #2})_{#3}}}
\newcommand\inexactd{d\hspace{-0.9ex}^-\!}

\newcommand\todo[1]{\textcolor{red}{(todo: #1)}}
\newcommand\fixme[1]{{\color{red}{(fixme: #1)}}}
\newcommand\wording[2][red]{\textcolor{#1}{\underline{#2}}}
\newcommand\wordingnote[3][red]{\textcolor{#1}{\underline{#2}\footnote{#3}}}
\newcommand\tyes{$\bullet$}

\newcommand\mypar[1]{\subsubsection{#1}  }
\newcommand\physicists{\mypar{Physicists}}
\newcommand\engineers{\mypar{Engineers}}
\newcommand\mathematicians{\mypar{Mathematicians}}

\newcommand\wrapme[1]{
  \begin{minipage}{.11\textwidth}
    \vspace{0.25em}
    \centering #1\\ \vspace{0.5em}
  \end{minipage}
}
\newcommand\tno{$-$}

\definecolor{small}{rgb}{0, 0, 1}

\newcommand\dbar{{\mkern6mu\mathchar'26\mkern-12mu d}}

\newcounter{excerpt}
\newcounter{excerptpart}[excerpt]
\renewcommand\theexcerpt{\arabic{excerpt}}
\renewcommand\theexcerptpart{\arabic{excerpt}\Alph{excerptpart}}

\newenvironment{excerpt}[1]%
  {\needspace{3\baselineskip}\refstepcounter{excerpt}\begin{itemize}%
    \item[]\hfill \textsc{Excerpt \arabic{excerpt}} \hfill #1
    \setlength{\itemsep}{0pt}%
    \setlength{\parskip}{0pt}%
  }%
  {\end{itemize}}

\newenvironment{excerptpart}[1]%
  {\needspace{3\baselineskip}\stepcounter{excerptpart}\begin{itemize}%
    \setlength{\itemsep}{0pt}%
    \setlength{\parskip}{0pt}%
    \item[]\hfill #1
  }%
  {\end{itemize}}

\begin{abstract}
  Partial derivatives are used in a variety of different ways within
  physics.  Most notably, thermodynamics uses partial derivatives in
  ways that students often find confusing.  As part of a collaboration
  with mathematics faculty, we are at the beginning of a study of the
  teaching of partial derivatives, a goal of better aligning the
  teaching of multivariable calculus with the needs of students in
  STEM disciplines.
  As a part of this project, we have performed a pilot study of expert
  understanding of partial derivatives across three disciplines:
  physics, engineering and mathematics.  Our interviews made use of
  the Partial Derivative Machine (PDM), which is a mechanical system
  featuring four observable and controllable properties, of which any
  two are independent.  Using the PDM, we probed expert understanding
  of partial derivatives in an experimental context in which there is
  not a known functional form.
  Through these three interviews, we found that the mathematicians
  exhibited a striking difference in their understanding of
  derivatives relative to the other groups.
  The physicists and engineers were quick to use measurements to find
  a numeric approximation for a derivative.  In contrast, the
  mathematicians repeatedly returned to speculation as to the
  functional form, and although they were comfortable drawing
  qualitative conclusions about the system from measurements, were
  reluctant to approximate the derivative through measurement.
  This pilot study led us to further questions.
  On a theoretical front, we found that existing frameworks for the
  concept of derivative are inadequate when applied to numerical
  approximation, and intend to address this with an expansion of the
  framework of Zandieh (2000).
%
%
  How do fields differ in their experts' concept image of
  \emph{partial} derivatives?
  What representations of partial derivatives are preferred by
  experts?
  We plan to address these questions by means of further interviews
  with a wider range of disciplinary experts.
\end{abstract}


\maketitle

\section{Introduction}
\label{intro}

Thermo is hard.  In a recent national workshop on the upper-division
physics curriculum, approximately $\nicefrac13$ of the faculty
indicated, in an informal show of hands, that they are uncomfortable
enough with the content of thermodynamics that they would be reluctant
to teach it.  There are a number of reasons \emph{why} thermodynamics
is hard.  One reason is the large number of different kinds of partial
derivative manipulations that need to be performed to solve many
theoretical problems.  In previous work, our group has analyzed expert
problem solving in this context using a framework of epistemic
games~\cite{kustusch2013expert, kustusch2014partial}.  Other research
groups have identified a variety of other difficulties both
mathematical and conceptual~\cite{bucy2007student, meltzer2009,
  christensen2010, thompson2012representations}.

In this paper, we are beginning an investigation of two further
reasons why thermodynamics is hard.  First, the nature of the
variables in thermodynamics.  The independent variables are measurable
(and changeable) physical variables such as pressure and volume rather
than immutable background markers such as space and
time.
\footnote{Yes, we know about relativity!} Furthermore, which of
these variables are independent and which are dependent varies with
the context.  In particular, the conjugate pair associated with
heating, i.e.\ temperature and entropy is known to be troublesome for
students~\cite{christensen2009student}.

Second, many important physical quantities in thermodynamics
are actually partial derivatives of other physical
quantities.  Thermodynamics involves an apparent surfeit of variables
in the sense that extensive variables such as volume have intensive
conjugate pairings such as pressure that have independent operational
definitions and are independently measurable, and may seem to be
independently controllable.
Because of this apparent surfeit of variables, thermodynamics is typically the
first time that physics students encounter scenarios in which the
quantities held fixed when taking a partial derivative are ambiguous.
In mathematics courses, students are taught that when taking partial
derivatives, all the \emph{independent} variables are held fixed%
.  Nevertheless, we have found that most
students come into our course with a firm belief that when taking a
partial derivative \emph{everything else is held fixed}.

Two years ago, one of us (DR)
developed the partial derivatives machine (PDM), a simple mechanical
device of springs and pulleys as a classroom manipulative.  (See
Section~\ref{sec:pdm} for a more complete
description of the PDM.)  Classroom activities involving the PDM
exhibit many of the same features as experiments and calculations that
students encounter in thermodynamics.  All the same issues about
independent and dependent, extensive and intensive variables arise.
And the question of which variables to hold fixed also arises, but in a
somewhat simpler context in that the variables involved are concrete
and tangible (lengths and forces).  Our hope was that students would
benefit from classroom experience with the PDM, and, to some extent,
this appears to be the case.

However, we have now become aware of a more fundamental underlying
problem.  In the first experiment with the PDM, students were asked to
find a partial derivative from experimental data.  Anecdotally, it
became clear that many students did not immediately understand that a
derivative can be effectively approximated by the ratio of small
numerical differences.

We do not believe that the issues we have observed with partial
derivatives are limited to students.  Indeed, we hypothesize that many
of the issues we have observed are due to the ways in which different
disciplines use and think about derivatives and partial derivatives.  In
this study, we conducted small group interviews with experts in several
STEM disciplines.  By studying experts' thinking about derivatives and
partial derivatives, we hope to obtain a better benchmark for comparison
in the study of students' thinking about those same ideas.  These
interviews were most similar to a clinical interview except that the
group setting provided a means for participants to listen and respond to
each other's ideas, rather than just the interviewers'.  The purpose of
this paper is explore the research question \emph{in what ways do
disciplinary experts in physics, engineering and mathematics think about
partial derivatives?}.

In the course of gaining insight into this question, we describe how the
responses were similar and differed across disciplines and consider the
role and affordances of the partial derivative machine in the experts'
responses.  In the remainder of this paper, we will give a description
of the Partial Derivative Machine (PDM), describe the method we used to
study our research question and give the results of our analyses from
the expert interviews.  We close by considering both the pedagogical and
the research implications of our results for student thinking and
learning in similar contexts.

\section{Theoretical Grounding}
\label{theory}

\subsection{Concept Images and Concept Definitions}


Thompson~\cite{Thompson2013} argued that the development of coherent
meanings is at the heart of the mathematics that we want teachers to
teach and what we want students to learn.  He argued that meanings
reside in the minds of the person producing it and the person
interpreting it.  We hypothesized that we could study experts' meanings
by studying their images, definitions, and representations for a concept
and using these to model their meaning for an idea.  We rely on
Vinner's~\cite{vinner} language of concept images and concept
definitions as an orienting framework.  Vinner described the concept
image as ``the total cognitive structure that is associated with the
concept, which includes all the mental pictures and associated
properties and processes'' (p. 152).  A concept definition is a verbal
definition that accurately explains the concept in a non-circular way.
While we were primarily interested in experts' concept images, as
illustrated by our tasks and method, we also considered their concept
definitions that in some cases underpin those images.  We see both the
concept image and concept definition as a means to operationalize and
explore the meanings that experts had for derivatives.

Vinner's definition of concept image explicitly allows a particular
concept image to involve many properties and many mental pictures.
We believe that mathematicians, engineers, and physicists have
multifaceted and detailed concepts images for derivative.  However,
Browne~\cite{browne2001case} showed that middle-division physics
students did not necessarily move spontaneously between various facets
even when changing to a different facet might make solving a particular
problem easier for these relative novices.  Our own classroom experience
bears out this observation.  Therefore, one of our research question is
to explore which facets of the concept image of derivative are cued for
different content experts by an open-ended prompt involving numerical
data from the PDM.

\subsection{A framework for student understanding of derivatives}
\label{sec:zandieh}

The framework developed by Zandieh for student understanding of
derivative is a valuable tool in this work~\cite{zandieh}.  This
framework is aimed at mapping student concept images for
derivative at the level of first-year calculus.  It begins by breaking
the formal symbolic definition of the derivative into three
\emph{process-object layers.}
\begin{equation}
  f'(x) = \lim_{\Delta x\rightarrow 0} \frac{f(x+\Delta
    x)-f(x)}{\Delta x}
\end{equation}
These three layers are the \emph{ratio layer}, in which one finds a
ratio of changes, the \emph{limit layer} in which one takes the limit
as the changes become small, and finally the \emph{function layer}, in
which one recognizes that this could be done for any value of $x$, and
thus describes a function.  These three layers are each required in a
complete understanding of derivative.  Moreover, each of these layers
can be seen both as a \emph{process} and as a reified \emph{object}.
As a \emph{process}, each layer is a procedure that you could use to
find a value.  But alternatively, one can understand each of these
layers as a static \emph{object}, which exists independently, and can be
and is acted upon by other processes.

Zandieh identifies an orthogonal dimension of \emph{representation}
(or alternatively \emph{context}) with four possible representations:
graphical, verbal, symbolic, and ``paradigmatic physical.''  Each of
these representations exists for each process-object layer.  We
introduced Zandieh's symbolic representation in the previous paragraph
and will here briefly outline the graphical representation of
derivative, which is slope.  At the \emph{ratio layer}, the graphical
representation is the slope of the secant line to a curve (which
itself is the graphical representation of a function).  At the
\emph{limit layer}, one has the slope of a tangent line.  And finally,
at the \emph{function layer}, one recognizes that the slope of the
tangent line is itself a function that could be visualized as a
curve.

This framework is valuable because it makes explicit the three
process-object layers that exist in the concept of the derivative, and
which can be used separately.  In Section~\ref{sec:meanings} we will
introduce our perspective on the different representations of
the derivative, which is expanded beyond that considered by Zandieh in
order to explicitly include physical representations at a level beyond
that treated by Zandieh.

\section{Background and Literature}
\label{background}

The purpose of this section is threefold.  First, we describe in what
ways students and experts have been shown to think about the concepts
of derivative and partial derivative, and hone in on particular
difficulties for students.  Second, we articulate various meanings for
derivative and partial derivative that come from both research
literature and our own experience working with students and
colleagues.  In this subsection, we also briefly elucidate several
language issues that have arisen as we ourselves, from our different
disciplinary perspectives, have discussed these various concepts of
derivative and we detail the specific language choices that we have
made in this paper.  Finally, we consider the importance of studying
experts' thinking about derivatives and partial derivatives as a means
to identify important learning goals for students in physics,
mathematics and engineering.  The overarching purpose of the section is
to demonstrate that while mathematics and physics education research
have gained insight into students' thinking about derivatives, they
have not fully explored thinking about partial derivatives.
Understanding how experts think about these ideas is a natural first
step to exploring how we might want students to reason about them.

\subsection{Students' Ways of Thinking about Derivative}

A number of researchers have identified difficulties students have in
thinking about rate of change of one variable functions.  These
difficulties range from students thinking about a graph as
representing its derivative, confounding average and instantaneous
rate of change~\cite{orton},
conceptualizing rate as the slope or steepness of a
graph~\cite{weberreps}, using statistical meanings for average to
interpret average rate of change~\cite{weberavg}, and inattention to how
fast quantities are changing with respect to one another.  Some students
conflate the average rate of change of a function with the average value
of a function and compute average rate of change by computing an
arithmetic mean and do not distinguish between the graph of the function
and the graph of the function's rate of change~\cite{cetin}.

Researchers have suggested that some of these difficulties might be
attributable to students not conceiving of rate of change as a
quotient.
For instance, students often discuss the rate of change as a slope but
do not speak of slope as a quotient (the change in a function's value
being so many times as large as the corresponding change in its
argument).  Instead, they talk about slope as the function's
steepness~\cite{zandieh}.  As another example, students often use a
tangent line and rely on visual judgments to sketch the derivative
function \cite{ferrini}.
In yet another study, students who were able to correctly rank the
slope at points on a graph were less able to find the sign of the
derivative at those points~\cite{christensen2012investigating}.
While this approach is not necessarily
problematic, thinking about sliding tangent lines does not necessitate
images of variation beyond visual judgments of steepness.  Zandieh also
argued that students can answer many standard calculus questions without
needing to think about functions as relationships between variables nor
needing to think about the rate of change of one quantity with respect
to another.  Students often respond to the directive ``find the
derivative of $g(x)$'' by acting on a symbolic expression using standard
rules for differentiation, without thinking about functions or rates of
change.  For both slope and symbolic expression issues, students might
not see the role of ratio and quotient in understanding rate and
derivative.  Indeed, researchers have documented that these issues keep
students from thinking about the derivative as a ratio of changes in
quantities, or interpreting that ratio as measuring how many times
larger one quantity is than the other over a particular, small interval
\cite{zandieh, carlsoncov, garsow}.

There is only limited literature that addresses how students think
about rates of change in the context of functions of two variables.
In a math setting, Yerushalmy~\cite{yerushalmy} provides an indication
of natural questions that might arise as students conceptualize rate
of change in multivariable settings.  She illustrated students'
struggles with how to think about dependence in a system with three
quantities and how to represent multiple quantities and their changes
in a single graph.  Her students struggled to describe the change in a
particular direction of a linear function of two variables.  Part of
the reason may have been that there are infinitely many directions in
which to move from a given point at a constant rate, yet, in general,
each direction yields a different slope.

In a physics setting, researchers have investigated mixed partial
derivatives and differentials in thermodynamics~\cite{bucy2007student,
  thompson2006assessing}.  They address students' ability to translate
back and forth between ``physical processes'' and partial derivatives,
and found that students were more able to go from a partial derivative
to a physical process than the other way around.  Similarly,
researchers investigating physical chemistry have found that students
need help in interpreting mathematics in a thermodynamics
context~\cite{becker2012students}.

It is clear that students do not necessarily reason about
derivatives and rates of change as we might hope.  Furthermore, very little
literature has considered how these issues extend to functions of more
than one variable, particularly how students might think about partial
derivatives.  This is surprising, given that many scenarios in
mathematics, physics and engineering require thinking about systems or
scenarios in which many variables may be changing simultaneously.

\subsection{Summary of Important Meanings for Derivative and Partial Derivatives}
\label{sec:meanings}

\subsubsection{What is a derivative?}  Researchers have identified the need for
students to be fluent in looking at the concept image of derivative
using multiple perspectives.  Derivatives are commonly described as
\emph{slopes}, as \emph{ratios of small changes}, as \emph{difference
quotients}, and as \emph{rates of change}, among others.  However, these
phrases are used in different ways in different disciplines, and by
different individuals within those disciplines.  For instance, the
description of \emph{slope} as ``rise over run'' could be either
numerical or symbolic, as well as graphical.  This complexity reflects
the multifaceted concept images of the derivative held by experts.

\newcommand\rot[1]{\rotatebox{90}{#1}}

\begin{table}
\begin{tabular}{|r|c|c|c|c|c|}
\hline
& \rot{\textbf{graphical}}
& \rot{\textbf{numerical}}
& \rot{\textbf{symbolic}}
& \rot{\textbf{verbal}}
& \rot{\textbf{experiment }}
\\
\hline
\textbf{slope} & \tyes &&&& \\
\hline
\textbf{ratio of small changes} && \tyes &&& \\
\hline
\textbf{difference quotient} &&& \tyes && \\
\hline
\textbf{rate of change} &&&& \tyes &\\
\hline
\textbf{name the experiment} &&&&& \tyes \\
\hline
\end{tabular}
\caption{Representations of derivatives.}
\label{dreps}
\end{table}

A common framework for multiple representations is the \emph{Rule of
Four} introduced by the Calculus Consortium (see Hughes Hallett \emph{et
al.}~\cite{HH98}), in which key concepts are presented
\emph{graphically}, \emph{numerically}, \emph{symbolically}, and
\emph{verbally}.  In recent work~\cite{roundy2014name}, we proposed
adding a fifth representation, based on \emph{experiment}.  In order
to be clear in our discussion of the different facets of expert
concept images of the derivative, we will associate each of the verbal
descriptions of derivatives given in the previous paragraph with just
one of these representations, as shown in Table~\ref{dreps}.  We have
based this association on our interpretation of the technical usage of
these common descriptions within the mathematics community.  The usage
may be unfamiliar to our physics readers, but we hope that this
unfamiliarity will help the reader notice the nuanced differences in
language.  In particular, both the use of \textit{rate of change} to
refer to rates that do \emph{not} involve time and the technical term
\textit{difference quotient} are largely limited to mathematics.  We
will use the conventions indicated in Table~\ref{dreps} throughout
this paper, except where explicitly stated otherwise.

We therefore consider five different ways to understand and think
about the concept image of derivative, each of which is useful in
different scenarios.
\begin{enumerate}
\item
The \emph{slope} of the tangent line to a curve.  This representation
describes the slope as a geometric measure of the steepness or slant
of a graph of a function.
\item
A numerical \emph{ratio of small changes}, by which we mean an
explicit numerical quotient, involving actual values of the ``rise''
and the ``run''.  By ``small changes'' we mean small enough that the
quotient represents a reasonable estimate of the derivative (within
the physical context of the problem).
\item
The result of algebraic manipulation of a symbolic expression.
Formally, these manipulations involve the \emph{difference quotient}
\(\frac{f(x+\Delta x)-f(x)}{\Delta x}\), but in practice a memorized
set of derivative rules is used instead.  (For the purposes of this
paper, we will conflate these two algebraic manipulations.)  This
meaning for derivative is a process applied to an algebraic object,
but does not include an image of that process as measuring how one
quantity changes with respect to another.
\item
The \emph{rate of change} of one quantity with respect to another,
which is here used to mean a description in \emph{words} of that
change.  (Outside of mathematics, the word ``rate'' implies that the
second quantity is assumed to be time, but we will use this term more
generically.)  In this representation, the derivative measures
covariation, i.e. how one physical quantity changes with respect to
another.
\item
Finally, as introduced in our previous work~\cite{roundy2014name},
particular derivatives can be associated with particular experiments,
such as measuring the change of volume in a piston of gas as weights
are added to the top of the piston.  Determining which experiment
might correspond to which derivative provides a representation of the
derivative, which we call \emph{name the experiment}.
\end{enumerate}
In the second case, it is most natural to think of the derivative as a
\emph{number}.  One picks a point at which to take the derivative and
computes a numerical ratio.  While that number will be different at
other points---making the derivative actually a function---this aspect
of the derivative may often be ignored.  When considering the slope of
the tangent line to a curve, it is clearer that the derivative is a
function, but it is also natural to think of the derivative as a
number, the slope at a single point on the curve.  When using the
symbolic approach, the derivative is \emph{inherently} a function, and
while that function could be evaluated at a point, its value cannot be
determined until after its functional form is known.

Most of these aspects of the concept image of derivative (namely, 1, 3,
4, 5) find approximate analogues in Zandieh's framework~\cite{zandieh}.
Interestingly, the second aspect, which is not present in Zandieh's
framework, turned out to be the most important one in the analysis of
our interviews and is a major theme throughout this paper.

\subsubsection{What is a \emph{partial} derivative?}  Partial derivatives differ
from \emph{ordinary} derivatives in important ways.  How we understand
this difference can vary with how we understand derivatives.
\begin{enumerate}
\item
A tangent line turns into a tangent plane in three dimensions, and a
partial derivative becomes the slope of the plane \emph{in a given
direction} in the domain, at a given point.
\item
When considering a ratio of small changes, a \emph{partial} derivative
requires that we specify not only which quantities are changing, but also
which quantities to hold fixed.
\item
The algebraic procedure to find a partial derivative of a symbolic
expression is identical to that for an ordinary derivative, provided
there are no interdependencies among the variables in the expression.
\item
The verbal description of derivatives as rates of change must
explicitly mention the independent variable(s) in order to describe a
partial derivative.  e.g. the derivative of volume with respect to
pressure would be ``the rate of change of volume as pressure is
changed, with either temperature or entropy held constant.''
\item
The representation of derivatives in terms of experiments is designed
precisely to take into account which physical quantities are
controlled, and which are not.  As such, it is particularly well suited
to descriptions of partial derivatives.
\end{enumerate}
In thermal physics, and other areas of mathematics, the quantities
that are being held fixed are context-dependent.  In general, one has
a set of interrelated variables, of which a few may be fixed.  The
number of independent variables is itself context-dependent, and in
physical situations we are seldom provided with symbolic equations
connecting the set of interdependent variables.  More often we rely on
physical intuition and argumentation to establish how many variables
may be controlled independently.  ``If I fix the pressure, temperature
and number of molecules, I could measure the volume and the mass,
therefore I believe I have three independent degrees of freedom.''

How we respond to the ambiguity provided by abundant physical
variables depends deeply on our concept of a derivative.  However, if
students' concept of derivative is not rooted in an image of measuring how
fast one quantity changes with respect to one or more other
quantities, then it is unlikely they will understand derivative in
the ways we intend.


\subsection{The Need to Study Experts from Mathematics, Engineering and Physics}
\label{experts}

In the previous two sections, we have made the case that there are a
variety of ways to think about derivatives and partial derivatives, but
students have difficulty thinking in the ways we might intend because of
their inability to think about a derivative as measuring the ratio of
small numerical changes between quantities.  This is the case even for
functions of a single variable.  Earlier, we also argued that most
real-world scenarios involve reasoning about multiple quantities and the
relationships between them, a goal that seems especially problematic for
students who have difficulty reasoning even about simple systems.  Our
reasons for studying the thinking of a variety of experts across
mathematics, engineering and physics were a) we believed they would be
accustomed to working with situations involving multiple quantities and
relationships, b) we believed their experience would allow us to observe
sophisticated reasoning patterns that we could only hope to observe in
extremely advanced students, c) we anticipated their ways of thinking
about partial derivatives could help us identify ``end goals'' for how
we want students to think and d) we expected their thinking would vary
across disciplinary areas, allowing us to better understand how students
in these fields might need to reason in different ways about derivatives
and partial derivatives.

As mentioned earlier, the overarching purpose of the section was to
demonstrate that while mathematics and physics education research have
gained insight into students' thinking about derivatives, they have
not fully explored thinking about partial derivatives.  Furthermore,
understanding how experts think about these ideas is a natural first
step to exploring how we might want students to reason about them.  In
the subsequent sections, we describe how we studied experts' thinking
about these ideas.

\section{Method}\label{method}

\begin{figure}
\includegraphics[width=0.77\columnwidth]{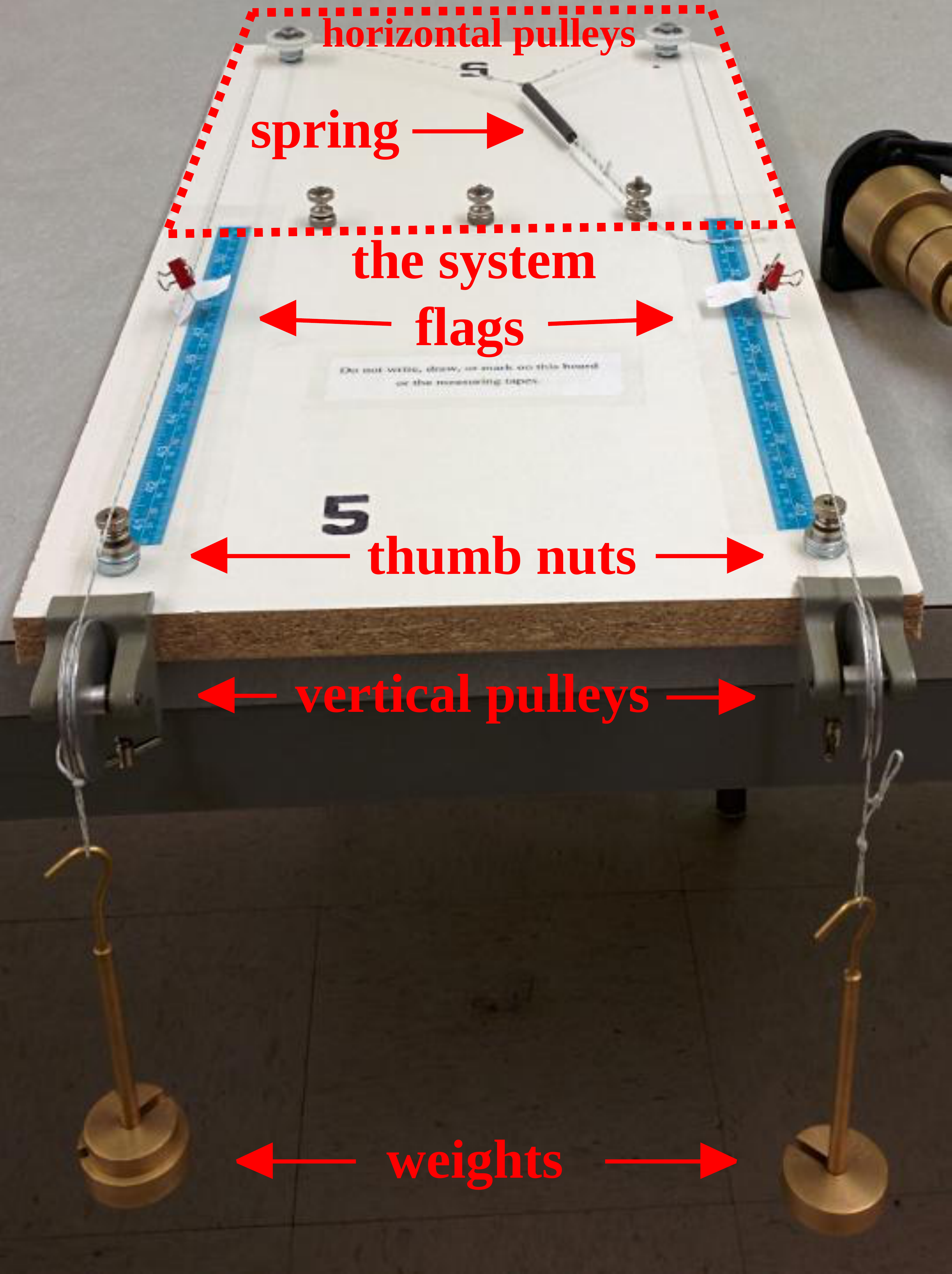}
\caption{The Partial Derivative Machine.  The machine consists of a
system (inside the dotted quadrilateral) which may be manipulated by
pulling on two strings.  Each string features a flag, which may be used
to measure its position, and a weight that can fix the tension in the
string.  In addition, thumb nuts may be used to fix the position of each
string independently.}
\label{fig:pdm}
\end{figure}

\subsection{The Partial Derivative Machine}\label{sec:pdm}

We have developed and used two versions of the Partial Derivative
Machine (PDM).  The first version of this device is documented
in reference~\citenum{sherer2013pdm}, and features a central system that is
attached to four strings.  The simplified version of this
device---which will be discussed in this paper---is shown in
Fig.~\ref{fig:pdm}, and consists of an anchored elastic system, which is
constructed of springs and strings.  In both versions, the elastic
system may be manipulated using two strings independently.  Each of
these two strings has a scalar position that can be measured with a
measuring tape and a tension that can be adjusted by adding to or
removing weights from a hanger.
Detailed instructions for constructing a Partial Derivative Machine,
including a parts list and photographs of additional central systems,
are available on our Paradigms website~\cite{portfoliospdm}.

The usefulness of the PDM emerges because it is an exact mechanical
analogue for a thermodynamic system.  The system contains a potential
energy $U$ (analogous to the internal energy) that cannot be directly
measured.  The system has four directly measurable---and
controllable---state properties: two positions $x$ and $y$ and two
tensions $F_x$ and $F_y$.  These four state properties play roles
analogous to volume, entropy, pressure and temperature in a
thermodynamic system.

Although the PDM has \emph{four} measurable and manipulable
properties, like its analogous thermal system, it only has \emph{two}
degrees of freedom.  One cannot independently control the tension and
position of a single string, unless one uses the other string to do
so.  In the PDM, the two degrees of freedom are physically manifest:
each corresponds to one string.  We can choose to manipulate that
string either by changing its position, or by changing its tension.

As in thermodynamics, the choice of which properties to treat as
independent variables is context-dependent.  While it is
experimentally easiest to control the two weights as independent
variables while measuring positions, it is sometimes theoretically
more convenient to view the positions as the independent variables.
Most notably, when using work to determine the potential energy, the
positions are the ``natural'' variables, as seen in the total
differential that is analogous to the thermodynamic identity:
\begin{align}
dU &= F_x dx + F_y dy
\end{align}
We can relate this total differential to the mathematical expression
\begin{align}
dU = \myderiv{U}{x}{y}dx + \myderiv{U}{y}{x}dy.
\end{align}
By equating coefficients of $dx$ and $dy$, we can find expressions for
the two tensions as partial derivatives of the potential energy.  
\begin{align}
F_x=\myderiv{U}{x}{y}  \qquad\qquad F_y=\myderiv{U}{y}{x}
\end{align}
This enables us to clarify the interdependence of the four directly
observable quantities.

\begin{figure}
 \includegraphics[width=0.6\columnwidth]{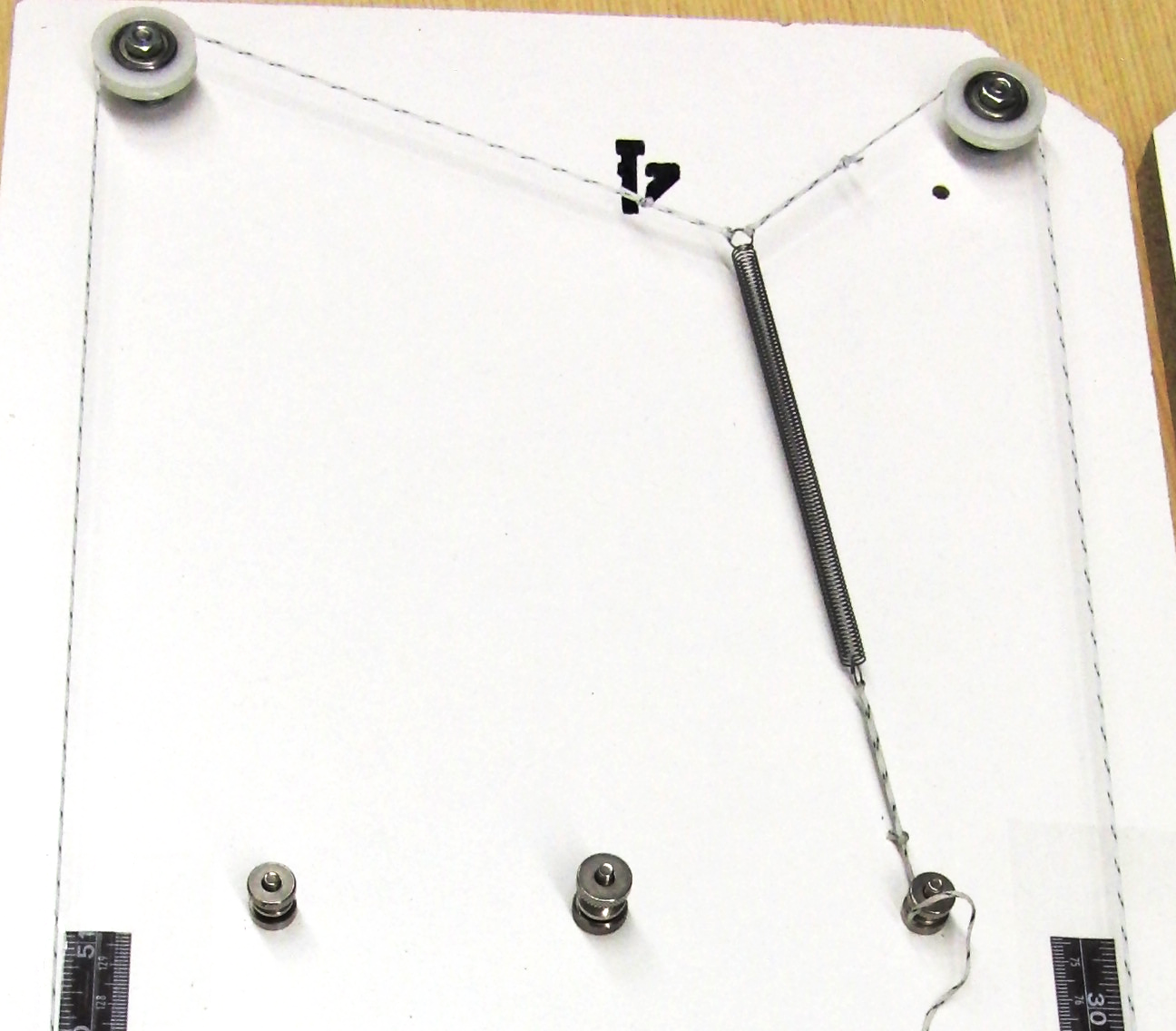}
 \caption{The system we used in the PDM for for the expert interviews.
 The system contains a single spring, attached by string to an
 off-center post.  The ``observable'' strings are routed around two
 horizontal pulleys towards the front of the machine where they are
 measured, and are attached to weights.}
\label{fig:system}
\end{figure}

We use the PDM to teach a mathematical introduction to thermodynamics
prior to our junior-level course in thermal physics, \emph{Energy and
Entropy}.  This introduction uses seven contact hours, and covers the
range of mathematical topics generally taught in undergraduate
thermodynamics.  We begin with total differentials and integration along
a path, discuss partial derivatives and chain rules, mixed partial
derivatives and Maxwell relations, and end with Legendre
transformations.  Throughout both this mathematical introduction and
\emph{Energy and Entropy}, a focus is placed on connecting the
mathematical expressions with tangible
reality~\cite{manogue2013tangible}.

\subsection{Design and Conduct of the Expert interviews}

To gain insight into our research question, we performed three expert
interviews, each of which lasted around an hour.  We interviewed three
groups of experts: physicists, engineers and mathematicians.  For the
\emph{physicists} we interviewed one associate professor and one full
professor.  Both use computational methods in their research, which is
in astrophysics and optics.  We interviewed three \emph{engineers}:
one chemical engineer who is a full professor with considerable
research and teaching expertise in thermodynamics, and two assistant professors
who study student thinking and epistemology in engineering.  Finally,
we interviewed two \emph{mathematicians} who are both assistant
professors and whose research is in mathematics education at the
collegiate level.  All of those we interviewed have doctoral degrees in
their disciplinary areas and work in physics, engineering, or
mathematics departments.

These interviews were mostly closely aligned with the notion of a
clinical interview in which one focuses on gaining insight into
another's thinking through systematic questioning and open ended
prompts.  At the same time, it is atypical that clinical interviews
occur in group format because of the difficulty of ascertaining an
individual's thinking.  Our use of groups was purposeful.  We believed
that the conversations between the experts and their questioning of
each others' responses would be just as important as the questioning
and prompts from the interview team.  Indeed, we saw that some of the
most interesting data we collected came from the experts debating each
others' responses.

\subsection{Prompts and Purpose of Task}

Our overall research question is to explore \textit{in what ways do
disciplinary experts in physics, engineering, and mathematics think
about partial derivatives.}  Important subquestions included how the
experts use notation, how they think about derivatives and partial
derivatives, how they think about the various variables and which, if
any, should be held fixed, and how they related those ways of thinking
to the PDM.  We chose to exploit a pedagogical design feature of the PDM
to explore our experts' understandings of partial derivatives outside
the world of symbolic manipulation.

We began by introducing the experts to the PDM, and showing them how to
manipulate the machine.   We then gave the experts the following prompt:
\[
\textit{Find } \frac{\partial x}{\partial F_x}\textit{.}
\]
We asked experts to ``find'' a partial derivative for which they are
given no functional form.  The PDM allows for measurement of changes in
the positions and tensions at discrete data points and the prompt was
designed in such a way that a ratio of changes in quantities would be
the only easy response.  However, we deliberately asked the prompt in an
open ended way so that it would not cue a particular aspect of the
concept of derivatives.  All parts of the system were chosen to be
visible in order to encourage discussion of the possibility of finding
an analytic expression by which one could determine the derivative,
although the actual determination of such an analytical expression would
have been prohibitively difficult.

When we provided this prompt, we did not define either $x$ or $F_x$,
but rather let the interviewees discuss what these quantities might
mean.  After they had discussed the meanings of these terms, and we
agreed that their meaning was sufficiently clear to us, we clarified
if necessary that $x$ was the position of one flag (i.e.\ one string)
and that $F_x$ was the tension in that string, which was determined by
the weight.  We note that this aspect of the task came at the
cost of not explicitly exploring graphical representations of tabular
experimental data with all groups of participants.

\subsection{Analytical Method}

The analysis of the data collected for this study relied on
systematically creating descriptions of experts' thinking about rate
of change (i.e. their concept images and definitions).  These
hypotheses resulted from in-the-moment observations and short
reflections by the researchers between interviews.  We also created
descriptions of experts' thinking about derivatives and partial
derivatives from the interviews using retrospective analysis.  The
retrospective analyses involved making interpretations and hypotheses
about participants' thinking by analyzing videos of the interviews
after they were completed.  Thus, we worked from two sets of
observations: those formed during the interviews and those formed from
analysis of the videos as a whole.  These analyses helped us to think
about the categories of concept images and definitions we present in
the analyses.  Thus, analysis of the data led to two things: a log of
the hypotheses developed about experts' thinking in the interviews and
descriptions of the experts' thinking made by the researchers after
the interviews were completed.  This data corpus provided a means to
describe patterns in experts' thinking, which in turn helped us focus
on various concept images and definitions they appeared to have for
derivative.

\section{Results}
\label{res}

During the course of the three interviews, a number of themes emerged,
each of which provided insight into the experts' concept images and
definitions for derivatives and partial derivatives.  These themes
were a combination of issues we noticed as we did the interviews and
issues that emerged as we conducted the data
analysis described in the analytical method section.  In the sections
below we describe these themes and articulate how we saw each group of
interviewees in the context of that theme.  Where it is possible, we
provide transcript excerpts from the interviews to support the claims
we make.

\subsection{Identifying $x$ and $F_x$}

Since the invention of the first partial derivative machine, we have
enjoyed sharing this tangible experimental system with colleagues.
During these informal ``interviews,'' we had noticed that different
individuals interpreted the symbols in the algebraic expression
\begin{equation*}
\frac{\partial x}{\partial F_x}
\end{equation*}
differently and began to expect that there might be disciplinary reasons
for these differences.  Therefore, we chose, in these interviews,
\emph{not} to tell the interviewees our own meaning for the symbols (at
first) but rather to let them explore in their groups what these symbols
might mean.  All three groups spent significant time thinking about and
debating over how to identify the quantities $x$ and $F_x$.  We expected
that the physicists and probably the engineers would share our
understanding that $x$ was the position of the pointer on the $x$
measuring tape and that $F_x$ was the tension in the $x$ labeled spring.
We expected that the mathematicians would try to invoke the mathematics
convention that a subscript indicates a partial derivative.  We were
surprised by several other unexpected types of confusion that our
notation caused.

The physicists initially identified $x$ with the elongation of the
spring and then wanted $F_x$ to represent a force in the same
direction as the spring. Their mental focus was clearly on the
internal mechanics of the system and they (at first) ignored the
measuring tapes on the PDM as a potential method for measuring a
position $x$.  When they recognized that their interpretation of the
symbols would lead to a ``total'' derivative rather than a partial
derivative, they then switched to an interpretation in which they wanted
the two strings to be perpendicular to each other, even going so far
as to manipulate the PDM to make this true.  Presumably, they were
invoking a physics convention that $x$ (and $y$) are independent
rectangular coordinates and $F_x$ (and $F_y$) are perpendicular
rectangular components of a single, net force vector.  We might
have anticipated some of this interpretation since one of the
interviewers has such a strong connection of the symbol $F_x$ to the
$x$ component of a force that she even referred to $F_x$ in this way in
the interview with the mathematicians.

The engineers immediately showed a preference for having $x$ represent
the horizontal coordinate (the strings were vertical) but acknowledged,
with laughter, that this must not be the case because it would have
trivialized the problem.  Unlike the physicists, they rapidly make use
of the left and right measuring tapes and are happy to invent their own
notation, calling the position of one pointer $x_L$ and the position of
the other pointer $x_R$.  They do not have any difficulty identifying
$F_x$ as the weight on one of the strings.

The mathematicians were not only puzzled by the meaning of the
subscript, as we expected, but also expected the capital symbol
$F$ to represent a function rather than a force.


\physicists The physicists began the task thinking that the position $x$
might be the elongation of the spring and $F_x$ a force in the direction
of the spring.  They rapidly corrected this one-dimensional
interpretation when they realized that they were asked to find a partial
derivative.  Excerpt~\ref{exc:phy-notation}.
\begin{excerpt}{5:00}
\label{exc:phy-notation}
\item[P2:]  So we have, probably it's the elongation of the of the
  spring, in this case.  So, right, so this is tied here [points to
  spring's anchored string], so if we put two forces like this [pulls on
  both strings] we need to, so, $x$, if this is $x$ [points to spring],
  right, we need to make the forces go on this direction [points in
  direction along spring].  So... 
\item[P1:]  But then it's a partial derivative, right? 
\item[P2:]  Right. 
\item[P1:]  So my feeling is that it cannot be that simple, or it
  would be a total derivative. 
\end{excerpt}  
They continued discussing the possible meanings of $x$ and $\vec{F}$ and
moved rapidly to language about \textit{balance (of forces)} and
\textit{perpendicular}. Notice the agreement in
Excerpt~\ref{exc:phy-notation2} between the two
interviewees.

\begin{excerpt}{5:57}
\label{exc:phy-notation2}
\item[P2]: Or if we, if we... Okay so this is one way [pulls on
right-hand string] or the other way is if we make it like this [pointing
to spring], right, so if there's a force, if we can somehow balance it
like this [pointing to spring and strings] and there's a force [pointing
in the direction of the spring] going this way because now this is
perpendicular [pointing to right-hand string where it is attached to the
spring], right? So this force will not count in $x$ [still gesturing
around the spring]. Then if we're able to balance it with the right
weights, then, you know, if we do like this [pulling on left-hand
string], right, maybe, you know the force that comes from this side. So
here we need to put a force that makes it [pulling on right-hand
string], right.

\item[P1]: At a right angle.

\item[P2]: Huh.

\item[P1]: That sounds like a good idea.

\item[P2]: Mhmm, and then, you know, then in this case, right, then
[pulling on weights to make a right angle] all the force on this [points in
direction of the spring] way is parallel to $x$, so that would be $x$ in this
case.

\item[P1]: Yeah, so then F x would be just this weight [points at
$F_x$].

\item[P2]: In this case, right.

\item[P2]: In the other case we have to do, have to analyze in two axes
as you said, and then figure out, you know, one of the two, so you have to
find the force on one of the axes and the... So, so so, the idea is to, we
need either to analyze $x$ [points at spring], right, in two directions and then
we'll have a net force, right, and analyze it into components in the two axes
or we just make it into one axis and then we have one force parallel to it,
right?

\item[P1]: I like your idea. I say if we put some weight at a right
angle [points toward string and spring angle at top of PDM], then we change
both weights [points at both weights] in order to keep the right
angle, and we measure displacement [points at measuring tape on left side] and
the force and we have...
\end{excerpt}

%
%
%
%
%
%
%
%
%
%
%
%
%
%
At 7:46, the
interviewer clarified the meanings of $x$, $y$, $F_x$, and $F_y$.
From this stage on, the physicists had no difficulty using this notation.

At the end of this portion of the interview, after the interviewers
interpretation of the symbols has been explained, P1 volunteers the
interpretation that, at the beginning of the interview he was trying to
measure the spring constant, which explains why he was looking at the
internal mechanics of the system.

\begin{excerpt}{8:37}
\label{exc:phy-notation3}
\item[P1] Yeah, I was more thinking we wanted to measure the spring
constant. 
\end{excerpt}

\engineers
The engineers began with the professional but humorous recognition
that there is an interpretation of the symbols that would make the
problem trivial:

\begin{excerpt}{3:09}
\item[E3]: I think it's easy if $x$ is this way [gestures perpendicular to
the strings].

\item[E1]: Yeah, exactly!

[Laughter]

\item[E3]: We don't know anything about that one, do we? [Laughter] So
we're done.
\end{excerpt}
They then gave a quick interpretation in words which seems, like the
physicists' interview, to focus on the internal mechanics of the system,
\begin{excerpt}{3:25}
\item[E3]: Okay, $x$ is going to be the spring.

\item[E1]: Sounds good.

\item[E2]: Okay so... $\partial x/\partial F_x$ [pronounced $d$ $x$
  $d$ $F$ $x$] is how much it moves per unit force, sort
of, could we do it that way?
\end{excerpt}
But within 1/2 minute, their attention focused on the strings and weights:
\begin{excerpt}{4:02}
\item[E3]: Does this [points to the written interview prompt 
$\frac{\partial x}{\partial F_x}$] mean
F in the direction, the x direction, only?
\item[E1]: So $F_x$ [pronounced $F$ of $x$], so I think it's a force in the x direction...
which the question is, does it matter [points back and forth to each string]
which side you're on?
\end{excerpt}
%
%
%
%
%
%
%
%
%
They then spent 2--3 minutes discussing how they expected the system to
behave as they added weights, based on the geometry of the system,
especially the angle of the spring.  During this analysis, they
discussed whether these symbols should refer to one side of the PDM or
both:
%
%
%
%
%
%
\begin{excerpt}{5:48}
\item[E2]: Do we want to go net, or do we want to pick one to
privilege? Obviously the right one is the real one and the left is
fake.

\item[E1]: So we could do each one separately

\item[E3]: Mhmm.

\item[E1]: And then verify that... that looks correct.

\item[E3]: Mhmm.

\item[E1]: In principle you could do one and figure out how the other
one...

\item[E3]: So it would be $\partial x/\partial F_x$ for this one [points to x string] and
then figure that out and then do it for this one [points to y string] and then
maybe some verification by comparing those two?
\end{excerpt}
After they decided to take data and began to construct the outline of
a table, they discussed how to label the columns of the table.  This
discussion returned them to the meaning of the symbols.  At this
stage, they agreed to add \textit{left} and \textit{right} subscripts
to their symbols.

\begin{excerpt}{8:07}
\item[E1]: So I would say... I would propose we do an x left and an x
right for each... but wherever we put the, the weights.

\item[E2]: Okay so we're going to do two trials.

\item[E1]: Because they're going to move opposite [motions opposite $x$
and $y$ movement] directions, right?

\item[E3]: Okay. So this [referencing table of $wt_{left}$
$x_{left}$ $wt_{right}$ and $x_{right}$] would capture the thing that
you want to do?
\end{excerpt}

%
%
%
%
%
%

\mathematicians
From the initial moments of the interview, the two
mathematicians puzzled over the meaning of the subscript on $F_x$,
noting that ``we have not seen this type of notation before'' and ``it
looks like a derivative but we are unsure what the symbols are.''
While they attended to the position of the strings, they did not
interpret $F_x$ as the force related to $x$ until the interviewer
explicitly suggested this to them part way through the interview.  In
Excerpt~\ref{exc:mat-notation}, the mathematicians explain their
confusion with the notation and in doing so, reveal that they often
associate a derivative with the process of differentiating an
explicitly defined function.  This was a theme we observed throughout
the interview with the mathematicians, yet only observed briefly in
the other groups.
\begin{excerpt}{2:24}\label{exc:mat-notation}
\item[M1:] I'm not familiar with the notation $F_x$ [pronounced $F$
  sub $x$]. 
\item[M2:] Me neither. 
\item[M1:] So, should we talk about what we think that might mean?
\item[M2:] Well, so, it's a par...  So, usually what we say, well,
  we...  We, I'm saying, my experience has been like 
  $\frac{\partial x}{\partial y}$
  [pronounced $d$ $x$ $d$ $y$] , right?
  The partial derivative of $x$ with respect to the $y$, right? So,
  like $y$ is some function of $x$ or... what's that...  Or you might
  have a function like little $f$ of $x$, $y$, \[f(x,y)\] um, is equal to some
  function of $x$ and $y$, and so you take, you know potentially that
  would be the partial derivative of that function... 
\item[M1:] With re... like of either $x$ or $y$ with respect to that
  variable, yeah 
\item[M2:] Yeah 
\item[M1:] Um, okay yeah, so I would think similarly, yeah, like
  that...  That symbol is the partial of $x$ with respect to some
  function.  So I, do you think that big $F$ sub $x$ means that it's a
  function that has $x$? Has at least as one of its variables, at least
  one of its variables, or do you think it means something else?
\item[M2:] Uh, no, I think that's what it means, that, yeah...
\end{excerpt}
A few minutes later, the mathematicians began to focus on the meaning
of the subscripts and revealed how problematic the notation was for
them.  In particular, we think their uncertainty reflects
possible disciplinary differences in notation, an issue we anticipated
prior to the interviews.
\begin{excerpt}{4:53}\label{exc:mat-other}
\item[M2:] So what if, what if we also had like an $F_y$? So how would
  that... be different, you know.  So they're both capital $F$, so yeah,
  boy I wish I could remember the meaning of the subscript [M1 points to
  $F_x$ on whiteboard] like if that already, like the original functions
  both $F$, capital $F$ is a function of $x$ and $y$.  And then when we
  see this notation [pointing to $F_x$ and $F_y$ on whiteboard], $F$ sub
  $x$, $F$ sub $y$ that means you've done something to that original
  function of $x$ and $y$.  You know, like if there's, that means that's
  the derivative with respect to $x$ [points at $F_x$] and that's the
  derivative with respect to $y$ [points to $F_y$]. 
\end{excerpt}
At 5:34 the interviewer interrupted to clarify that $F$ means
``force'' and that the $x$ subscript indicates the component of the
force in the $x$ direction.  The mathematicians immediately grasped
that there could then be a $y$ and an $F_y$, although---like the
engineers---they did not guess that $y$ would be the position of the
other string and $F_y$ the tension on that string.

\subsection{What is a (partial) derivative?}

The prompt to find a partial derivative yielded an interesting picture
of how our experts understood the concept of derivative.  In the
end, all of our experts found an approach to measure the derivative
experimentally---which practically requires application of the
\emph{ratio} layer of Zandieh~\cite{zandieh} (see
Section~\ref{sec:zandieh}).  Each group immediately proceeded to
explore whether their changes were sufficiently small.  This reflects
a strong recognition of the \emph{limit} layer of
Zandieh~\cite{zandieh}, and a recognition that a measurement of the
derivative must account for this.  We saw a large difference between
disciplines in the prominence of the \emph{function} layer of
Zandieh~\cite{zandieh}, which we recognize as an exploration of the
dependence of the derivative on the forces $F_x$ and $F_y$.  And
finally, we saw a large difference in the degree of comfort with a
numerical approximation for the derivative.

The physicists and engineers were very comfortable with the derivative
as a number, and quickly computed this number as a ratio of small
changes.  In both cases, they took an additional measurement in order
to verify that their changes were sufficiently small that they were
within the linear regime.  Their comfort with a single numerical
answer for the derivative suggests to us that the physicists and
engineers were satisfied with a derivative that omits the function
layer of Zandieh~\cite{zandieh}.  Both groups did acknowledge and
discuss that there is a functional dependence of the derivative on the
force.

Our mathematicians, in contrast, saw the derivative as a function, and
expressed concern about numerical approximation.  Interestingly,
although the mathematicians were persistent in seeking a symbolic
functional form for the derivative, in the process they were quite
comfortable and creative with drawing conclusions about the derivative
through experimentation, and specifically investigated the functional
behavior of the derivative in how it changes when different parameters
are modified.

Unlike the physicists, the engineers went on to mention other
representations of the derivative in their discussions, such as the
slope of a graph, and a symbolic expression derived from statics.  The
mathematicians made use of essentially every representation for
derivative \emph{except} for the ratio of small changes.

\physicists After some technical difficulties, at 13:56 the physicists
had collected their first data, which was sufficient to find a simple
ratio of changes.  In Excerpt~\ref{exc:phy-find-ratio}, they discussed
whether two values each for $x$ and $F_x$ are sufficient.  In
particular, they expressed the idea that ``because it is a derivative''
it may need smaller increments.
\begin{excerpt}{14:29}\label{exc:phy-find-ratio}
\item[P2:] The other question is, because it is a derivative, does it
  need to have smaller increments? 
\item[P1:]  Umm... 
\item[P2:]  The difference is... 
\item[P1:]  Yeah... unless the system is linear or not. 
\item[P2:]  So maybe we should try... 
\item[P1:]  Try it with fifty. 
\end{excerpt}
The went ahead and took an additional data point to verify that their
$\Delta F_x$ value was small enough that they were working in the
linear regime, which they did in Excerpt~\ref{exc:physicists-linear}.
They computed the derivative as a ratio of their $\Delta x$ and
$\Delta F_x$, confirmed that the two sizes of $\Delta F_x$ gave
similar answers, and concluded that they had made a measurement of the
derivative.
\begin{excerpt}{15:48}\label{exc:physicists-linear}
\item[P1:] And, um, and now we have everything, we can do $\Delta x$
  over $\Delta T$ or $F$ [P1 had previously referred to the force as
    ``tension''].  And because we held constant $F_y$, it uses the
  partial derivative. 
\item[P2:]  Okay. 
\item[P1:] And that is 1.5 over one hundred [pointing to table of $x$
  and $F_x$ values] which is
  \[\frac{\Delta x}{\Delta F}=\frac{1.5}{100}\] 
\end{excerpt}
They proceeded to discuss the possibility that this derivative might
not be a constant, i.e. might depend on the value of $F_x$,
acknowledging that that the derivative is a function, although they
did not feel that this it was necessary to explore this dependence in
order to answer the prompt.

\engineers We begin with the engineers at the same stage as we began
with the physicists, after they had taken two measurements each of $x$
and $y$ ($x_{\textit{left}}$ and $x_{\textit{right}}$ in their
notation) with different values of $F_x$ and the same $F_y$, and thus
had enough data to compute an estimate of the derivative as a ratio of
changes.  In Excerpt~\ref{exc:eng-process} the engineers discussed
whether to go on taking more data.
\begin{excerpt}{14:17}
\item[E1:] So shall we put another hundred grams on to see if it's
  linear? 
\item[E2:] I think so.  Yeah. 
\item[E3:] I'm feeling like we should be writing some huge equation to
  describe this and not have to mess with this, but I'm unwilling to
  start that procedure. 
  \label{exc:eng-process}
\end{excerpt}
Also, in Excerpt~\ref{exc:eng-process}, E3 suggested that perhaps they ought
to be performing a symbolic calculation, but was reluctant to do so,
and the subject was dropped.  On another occasion, the engineers spent
some time discussing the possibility of using statics to find a
symbolic solution, and also dropped the idea.  They clearly recognized
the possibility of a symbolic expression, but were unwilling to pursue
it.

At this point the engineers computed values for the derivative for two
values of $F_x$ using a ratio of changes, and found somewhat different
numerical values.  In Excerpt~\ref{e:eng-slope} they discussed how to
interpret these differing ratios.
\begin{excerpt}{18:49}\label{e:eng-slope}
\item[E3:] Are those the same? 
\item[E1:] Well so, I would recommend, let's let's crank it a bunch
  and see if we come up with the same number. 
\item[E3:] Yeah. 
\item[E1:] And then if it's not, we could either, we could either plot
  it and take slopes, or we could say, hey is that good enough. 
\item[E3:] Right. 
\end{excerpt}
After some further discussion they concluded that the two slopes were
the same within their experimental error, and that they had a good
measurement of the derivative.  In checking that the response is
linear they addressed the \emph{limit layer} in a way that could
be surprising: they \emph{increased} the size of their change in order
to show that it was sufficiently small.  This expert behavior
reflected a recognition that experimental uncertainty would make
\emph{smaller} changes harder to measure.

The engineers returned to the idea of graphing and finding the slope
much later in the interview.  After being prompted with how they would
perform an analysis if they \emph{had} sufficient experimental data, the
engineers returned in Excerpt~\ref{exc:eng-graph} to the idea of the
slope of a graph to explain how they could find the $F_x$ dependence
of the partial derivative.
\begin{excerpt}{38:43}
  \label{exc:eng-graph}
\item[E2:] I would like to see it, to see them, like what you said
  about slope, so like that.  So what would the $x$... 
\item[E1:] So we could, I mean we could do a plot, right, of mass
  so... 
\item[E3:] Versus uhh... 
\item[E1:] Yeah, we would want on the $x$ axis.  Well we would want, um,
  force, right? [E3 begins making plot of $x$ and $F_x$] That's our
  independent variable and length or mass. 
\item[E3:] And wouldn't it be like delta, well it would be force no
  no, of course not. [Labels plot with $x$ on vertical axis and $F_x$ on
  the horizontal axis] $F_x$ and then this would be... 
\item[E1:] And that would be $x$. 
\item[E3:] $x$. 
\item[E1:] And then we could plot that and we could plot that... we
  could do a series of graphs set at different $L_2$'s. [points to the
  subscript $L_2$ on $\left(\frac{\partial x}{\partial F_x}\right)_{L_2}$]
\item[E3:] Uh huh. 
\item[E1:] And then just at any value, the slope of that would be this
  derivative. 
\item[E3:] Yeah, I like that, yeah, I do. 
\item[E2:] So if we wanted, say it was a a function, we know enough to
  figure out the function, right? 
\end{excerpt}
In this discussion the engineers gave a clear picture of how they
could graphically obtain the derivative as a function of both $F_x$
and $y$ (which they called $L_2$.  The engineers continued to further
discus how they could obtain the slope from the graphical data by
performing a curve fit.

\mathematicians The mathematicians spent much longer than the
physicists or engineers before finding an answer to the prompt that
they were satisfied with.  In the process, they used physical
manipulation of the machine to reach several conclusions regarding
properties of $x$.

Very early on, the mathematicians grappled with identifying the
arguments of the function $x$.  This was in strong contrast to the
physicists and engineers, who did not talk about $x$ as a function until
after having experimentally found the partial derivative.  In
Excerpt~\ref{exc:mat-x-of-two-forces}, the mathematicians used physical
reasoning to conclude that $x$ is a function of $F_x$ and $F_y$.  This
discussion followed considerable manipulation of the machine.
\begin{excerpt}{11:56}\label{exc:mat-x-of-two-forces}
\item[M2:] So, yeah, so $x$ is our position $x$ [points to $x$ position
  marker], like if we decide that $x$ is some position on here [points
  to measuring tape on $x$], um, that's going to be a function of the
  weight we have here [points to $F_x$ weight], right? 
\item[M1:] Um hmm. 
\item[M2:] It seems like $x$ has to be a function of something, it's
  not just...  It's not going to be constant, right? 
\item[M1:] Right, right. 
\item[M2:] It's going to be, it's going to depend, but doesn't it
  depend on both of these things [pulls on both $F_x$ and $F_y$
    weights], right? Because I can leave this constant here [lets go
    of $F_x$], but then this is gonna...  If I move, if I add weight
  here [pulls on $F_y$], then... 
\item[M1:] Right, right.  Then if we like tie this off [clamps $y$
  string], then maybe $x$ really does just depend on this guy [pulls
  $F_x$ weight], whereas if it's here [unclamps $y$ string] and on
  both... 
\end{excerpt}
The mathematicians have concluded that by fixing $y$, $x$ becomes
independent of $F_y$.  This reasoning probably reflects an
interpretation of $F_y$ as the mass on the hanger, rather than the
tension in the string---an interpretation that is entirely consistent
with the information they were given, although this was not the
interpretation we intended.

A few minutes later, in Excerpt~\ref{exc:mat-rates-of-change}, the mathematicians
discussed the derivative as ``rate of change'' and addressed how
their understanding of $x$ as a function of the two forces relates
to the partial derivative they were asked to find.
\begin{excerpt}{14:26}\label{exc:mat-rates-of-change}
\item[M2:] Well so [points to $\frac{\partial x}{\partial F_x}$ on
  whiteboard], um, so you know, think about derivatives as rates of
  change, right? So, if you think about like in this [$\frac{\partial
  x}{\partial F_x}$], how I usually look, you know, if we were looking
  at this lowercase $f$ of $x$, $y$ [points to $f(x,y)$], the derivative
  of $f$ with respect of $x$, that partial derivative is like the rate
  of change with respect to $x$.  But now $x$ is gonna be dependent on
  these two forces and so it's like we're finding out the rate of change
  in $x$ with regard to this particular force [points to $\frac{\partial
  x}{\partial F_x}$].  So originally that notation is like why, `these
  physicists are doing it all wrong, they're putting it all in the wrong
  place!'  But now I understand that, now I understand the
  notation.  Does that make sense? 
\item[M1:] Uh huh. 
\item[M2:] So if $x$ is dependent on these two forces [pulls on $F_x$
  and $F_y$], then, right, which we're figuring that out, um, then this
  notation makes sense. 
\item[M1:]  Mhmm.  Mhmm. 
  \label{exc:identify-this}
\end{excerpt}
After this, the mathematicians began discussing how to address the
prompt.  This led to a discussion of what would happen if they fixed
the $y$ string, which appears in Excerpt~\ref{exc:mat-clamping-y} in
the following section.

In the middle of the interview, the mathematicians began discussing
what it is that they are being asked.  In
Excerpt~\ref{exc:mat-assigned-task} the interviewer responded by
explicitly asking what they think their task is, and clarified that
they were not actually asked for a symbolic expression for the partial
derivative.
\begin{excerpt}{28:00}\label{exc:mat-assigned-task}
\item[INT:] So what do you think your assigned task is? 
\item[M2:] To find an expression that represents... [points to the
  $\partial x/\partial F_x$ on the board] 
\item[INT:] I never said to find an expression, I just said to
  find this derivative. 
\item[M2:] Find that derivative, okay. 
\item[M1:] And so...  I mean I guess this is a dumb question, but I'm
  asking it.  So like what's the nature of what we're trying to find,
  like is it a number, is it a function, is it a, an expression? 
\item[M2:] Right. 
\end{excerpt}
This discussion highlighted their confusion over what it was they were
being asked to find.  They continued with a discussion of how a
derivative could both have a numerical value and be a function at the
same time.  A few minutes later, they addressed this question
directly in Excerpt~\ref{exc:mat-velocity}, in which M2 remembers from
calculus texts how the position of a ball can both be a number and a
function.
\begin{excerpt}{43:55}\label{exc:mat-velocity}
\item[M2:] But if position is just like a number, like forty five or
  forty four, forty six or forty seven,  then any derivative of the
  numbers is just going to be zero. 
\item[M1:] Yeah, well on the position is, I mean, yeah, I see what
  you're saying. 
\item[M2:] But then, you know, our calculus text, our calculus texts
  talk about a position function where a ball is flying through the
  air and position is a quadratic function, so it's just two times
  something... Right? Um, or negative, depending on the coefficient of
  x squared, so... 
\end{excerpt}
Throughout the interview, the mathematicians returned to
speculation as to the functional form of $x$.  In some cases, they
used hypothetical symbolic expressions to reason about possible
behavior of $x$.
As the end of the time allocated for the interview approached, the
interviewer began pressing the mathematicians for an explicit answer
to the prompt.  In Excerpt~\ref{exc:mat-ordered-pairs} one
mathematician replied jokingly with a guess as to a symbolic expression.
When pressed by the interviewer, the mathematicians ended up
describing a process by which they could find a numeric answer from
the slope of a graph of their data.
\begin{excerpt}{44:54}\label{exc:mat-ordered-pairs}
\item[M2:] It's one over $x$! [Laughs] 
\item[INT:] Okay, so you don't know, you don't know a functional
  relationship. 
\item[M2:] No. 
\item[INT:] So, what else could you do? 
\item[M2:] I don't know...  Take a partial derivative to find that
  expression? [references $\partial x/\partial F_x$]...  Um, well I
  know that when, you know, like if I were just, if it was...  I mean
  to find the slope, right, between two points to approximate the
  derivative, yeah? 
\item[INT:] I'm giving you my blank interviewer face! [joking] 
\item[M2:] I know! [Laughter and inaudible joking] 
\item[M2:] Um, right so, if we didn't know the function and one way
  you approximate the tangent, right, or the slope at a particular
  point which is a rate of change, you just find two points close
  together and find slope.  So, if we find, if we know change in the
  force and the change in the $x$ [points to $\frac{\partial x}{\partial
  F_x}$], like we can take some of our ordered, we can consider these to
  be like ordered pairs [referencing table of values for $F_x$, $F_y$,
  $x$, and $y$] and then just approximate...  Uh, you know come up with a
  numerical value. 
\end{excerpt}
The mathematicians proceeded to write ordered pairs of numbers on
their board (using data that they had previously collected), rather
than drawing the graph they described.  Like the engineers and
physicists, soon after they recognized the existence of a numerical
solution, they began considering whether those changes were small
enough.  In Excerpt~\ref{exc:mat-small-enough}, the mathematicians
discussed the accuracy of their approximation, and concluded that they
could improve it by adding smaller weights.
\begin{excerpt}{47:30}\label{exc:mat-small-enough}
\item[M1:] Okay so we have those differences, and so these are, we can
  think of them as just being points on our function like whatever
  that function is. 
\item[M2:] Actually a really rough approximation...  
\item[M1:] An approximation.  And so, we can just sort of consider the
  slope of those. 
\item[M2:] Mhmm. 
\item[M1:] Right? And that would be a reasonable approximation. 
\item[M2:] Mhmm.  Well, I don't know about reasonable. [Laughs] 
\item[M1:] Yeah, but I mean that would be something that could
  approximate that... 
\item[M2:] Cause I mean you know you might want to think about getting
  [points at an ordered pair] more fine grained slope by like just
  increasing by, you know, what are these, grams? [looks at weights]
\end{excerpt}
After this, they spent some time adding small weights and measuring
changes in $x$---although they never did actually perform a division
or write down a ratio.

\subsection{What should be held fixed, and does it matter?}

A defining feature of a partial derivative is that some other
variables must be held fixed.  In thermodynamics in particular, this
choice of what to hold fixed has important implications, and we were
very interested to see how our experts treated this question.  We
believed that the prompt to
\begin{equation*}
  \textit{Find }\frac{\partial x}{\partial F_x}
\end{equation*}
was ambiguous, and were interested to see how our experts responded to
this ambiguity.  We expected that at least some experts would point
out the ambiguity and ask us which property to hold fixed.  Our
interview had other results: all our experts assumed that since we
were taking a partial derivative with respect to one force, then the
other force must be held constant.  In this respect, the PDM differs
from thermodynamics, in which a derivative such as
\begin{equation*}
  \frac{\partial p}{\partial V}
\end{equation*}
does not have any inherent information about whether entropy or
temperature should be held fixed.

In further discussion, each group addressed the question of what would
happen if they fixed $y$ instead of $F_y$.  Our experts felt that
clamping the $y$ string would change either the system or the function
$x$, and thus the derivative would change.  The mathematicians
measured how $x$ changed with and without $y$ fixed, and
concluded that it moved less when $y$ was fixed, although they did not
discuss this result in terms of derivatives.

Also related to this question is whether the derivative depends on the
value of $F_y$.  This is a further level of the concept of
\emph{function} than we explored in the previous section.  Ordinary
derivatives are a function of the single independent variable, but
partial derivatives are functions of multiple independent
variables.

\physicists One physicist decided that $F_y$ must be held fixed when
measuring ${\partial x}/{\partial F_x}$, because they were taking a
derivative with respect to $F_x$.  Thus if they took the inverse
derivative ${\partial F_x}/{\partial x}$, then $y$ would have been
held fixed.  He stated in Excerpt~\ref{exc:phy-other-deriv} that
this is ``what they taught me.'' When he considered fixing the thumb
nut, he stated that this (physical act?)  would change $x$ to be a
different function, a function of $F_x$ and $y$.

In Excerpt~\ref{exc:phy-which-held-fixed}, before starting to take
data the physicists began discussing the meaning of the partial
derivative, and what to hold fixed.
\begin{excerpt}{10:04}\label{exc:phy-which-held-fixed}
\item[P1:] Yeah, is $\partial x$ over $\partial F_x$? If it was
  $\partial F_x$ over $\partial x$, it would be different. 
\item[P2:]  Right. 
\item[P1:]  Because we are keeping it constant, two different
  things. 
\item[P2:]  Right. 
\item[P1:] So if it is $\partial x$ over $\partial F_x$, the dependent
  variable is $F_x$, so we need to keep constant $F_y$. 
\item[P2:]  Right. 
\item[P1:]  Which is simple, it means we assume there is weight on
  each side.  Then we completely ignore that part. 
\item[P2:]  Right.  What if we just clamp it here though? In this case
  there's no, there's no, there's no $F_y$ because there's no, I mean
  this is constant, right? So there's no.... 
\end{excerpt}
At this stage, they spend a minute clarifying which string was $x$ and which
was $y$.  Then in Excerpt~\ref{exc:phy-clamping}, they continued their
discussion of what would happen if they clamp the $y$ string.
\begin{excerpt}{11:17}\label{exc:phy-clamping}
\item[P2:]  Okay, right.  So we need $F_y$ to be constant, so should we
  clamp this [points to $y$ clamp]? 
\item[P1:]  Should we clamp...  No, I'm going to say no because if we
  clamp it [points to $y$ string], then we do not hold the tension
  constant.  If we clamp it, we keep $y$ constant, not $F_y$. 
\item[P2:]  Right. 
\item[P1:]  But we need to keep $F_y$ constant, so I would let it move
  and just make sure we don't change the weight here [points to
  $F_y$]. 
\label{exc:phy-which-fixed}
\end{excerpt}
Thus the physicists concluded that they should fix $F_y$ in order to
measure the partial derivative with which they were prompted.

Later in the interview, the interviewer prompted the physicists by
asking them to measure the derivative
\begin{equation*}
  \myderiv{x}{F_x}{y}.
\end{equation*}
In Excerpt~\ref{exc:phy-other-deriv}, the physicists correctly
explained how to measure this derivative by clamping the other string.
They also explained about why they had assumed that $F_y$ should be
held fixed, and went on to discuss how a partial derivative is like a
derivative along a path in a multidimensional space.  They end by
suggesting that clamping the $y$ string changes the system itself,
creating a different function.
\begin{excerpt}{24:27}\label{exc:phy-other-deriv}
\item[P1:]  Now we can clamp that [indicates $y$ string]. 
\item[P2:]  Okay...  So now... 
\item[P1:]  But that's the natural expectation if you don't specify
  what you're holding constant.  You're holding constant the
  other.  That's what they taught me. 
\item[INT:]  That's what... 
\item[P1:]  They taught me.  Like when I was a student. 
\item[INT:]  So tell me exactly, what exactly did they tell you
  when you were a student? 
\item[P1:]  That when you do a partial derivative of a function that
  is a function of, um, more than one independent variable, you take,
  you do the incremental [pointing to $\frac{\Delta x}{\Delta F}$ on
  whiteboard], keeping all the independent variables
  constant. 
\item[INT:]  Okay. 
\item[P1:]  Now if you're doing the second thing, it's real, funny
  because you are telling me to keep a constant, a function
  constant.  Because $y$ [points to $y$ string] is always a function of
  the two forces [points to weights]. 
\item[P2:]  Right, so it's... 
\item[P1:]  So it's not something that... 
\item[P2:]  So you take the derivative in a different direction in $F_y$
  and $F_x$ plane then you would take it if you were to keep $F_y$
  constant.  So the partial derivative, right, is in the specific
  direction, so it depends on the direction. 
\item[P1:]  It seems to me something like a partial derivative along a
  path, more than along an axis. 
\item[P2:]  Yeah, exactly.  So you do it in a different, along a
  different path or a different... 
\item[P1:]  Along a certain path that is not, I don't know what to
  call it. 
\item[P2:]  Right, yeah. 
\item[P1:]  Let's do it.  
\item[INT:]  But $y$ is not separately independent? 
\item[P1:] $y$... well you can if you want, I guess.  Say that, I mean
  the moment you clamp it [points at clamp on $y$ side], you change your
  system.  And you can say now $x$ is not a function of $F_x$ and $F_y$,
  it's a function of $F_x$ and $y$.  That is a different function.
\end{excerpt}

\engineers
At a point in the middle of the interview, E2 pointed out that they
had been instructed that they could fix $y$ using the clamp, leading
to a discussion of what that would mean and how it would affect their
results.
\begin{excerpt}{27:00}
\item[E2:] They said we could stop [inaudible, points at the thumb nut
  for fixing $y$.].  Does that help us at all? 
\item[E3:]  That would be crazy! [jokingly] 
\end{excerpt}
After this the engineers spent a couple minutes discussing what they
would find by performing the same derivative measurement while the
thumb screw holds $y$ fixed.  After a bit of discussion involving how
to perform the measurements, E1 concluded in
Excerpt~\ref{exc:eng-different-constant} that they are now holding
something different constant.
\begin{excerpt}{29:06}\label{exc:eng-different-constant}
\item[E1:]  We're holding something different constant, and we could
  think about, if, once we look at the numbers and see how they define
  that. 
\item[E3:]  But it's sort of like you're saying, certainly a different
  problem holding that [points to $y$ string] constant. 
\item[E1:]  Yep. 
\end{excerpt}
At this point the engineers proceeded to take some data with $y$
fixed.  Then the engineers discussed how to notate what they
just measured with $y$ fixed.  E1 (who teaches thermodynamics)
suggests that they use the subscript convention to notate which
quantity is held fixed.
\begin{excerpt}{35:58}
\item[E1:]  So that would be a partial derivative, I think that's what
  they asked for. 
\item[E3:]  So that way we want to write... 
\item[E1:] Partial of $x$, yeah, with respect to $F_x$ and then put
  parentheses around that whole big thing. 
\item[E3:]  This thing? 
\item[E1:]  Yeah...  And then write at the bottom,
  subscript $L_2$. 
\item[E3:]  I'm not very good at subscripts [jokingly], I've heard at this
  meeting earlier... [continuing earlier joke] 
  [E3 writes on board:]
  \[ \myderiv{x}{F_x}{L_2} \]
\item[E1:]  That's awesome. 
\end{excerpt}
The engineers at this stage were using the symbol $L_2$ for the
distance we call $y$.  Thus this notation is in complete accord with
our understanding of the problem.  We find this unsurprising, given
that E1 is entirely familiar with thermodynamics.  E3, who was writing
on the board, seemed considerably less comfortable with this notation.

\mathematicians We saw in Excerpt~\ref{exc:mat-x-of-two-forces} a
mathematician said that fixing $y$ would change the system by making
$x$ no longer a function of $F_y$.  This would make $x$ a function of
only one variable, since they did not talk about the value of $y$ as a
variable.
A few minutes later, in Excerpt~\ref{exc:mat-clamping-y}, the
mathematicians explored what happens if they fix $y$ by clamping its
string.  They observed a smaller change in $x$, and decided that this
made sense, since the clamp impedes the motion.
\begin{excerpt}{17:48}\label{exc:mat-clamping-y}
\item[M1:] But like if I... if we turn this off [clamps $y$]... 
\item[M2:] Oh yeah. 
\item[M1:] Does the same thing happen... [adds weight to $F_x$] So it
  moved a lot less, but, I mean... Okay, well, let's just see what we
  get, so this is like [takes $x$ measurement]
  a little over forty four. [a smaller change in $x$]
\item[M1:] Which makes sense because... Well, I mean, this [touches
  $y$ clamp] seems harder, like this is actually not allowing it to
  move at all. 
\end{excerpt}
After this, they spent some time discussing what value of $F_x$ to use
for their derivative, not being sure what to focus on.  We believe
based on the above quote that they understood this to mean that the
derivative itself was smaller if $y$ were fixed, although they did not
explicitly state this.

\subsection{How many independent variables are there?}

As discussed in Section~\ref{sec:meanings}, the number of independent
variables---or the number of degrees of freedom---is a critical
property of a physical system.  It determines the number of parameters
we must control in order to determine the state of that system, and at
the same time limits the number of parameters we can fix when finding
a partial derivative.  The PDM has two degrees of freedom, as
discussed in Section~\ref{sec:pdm}.

The existence of two degrees of freedom means that the derivative
$\partial x/\partial F_x$ (and the variable $x$ itself) should be
viewed as a function of two variables.  This is an expansion of the
layer of \emph{function} into multiple dimensions.  In this section,
we discuss how our experts interpreted the number of degrees of
freedom of the system.

The question of how many independent variables were present arose in
each interview.  Both the physicists and engineers treated this question
explicitly (the physicists, at the prompting of the interviewer), and
went through a stage of talking of $x$ being a function of the remaining
three variables.  They then concluded that one could eliminate one of
those three, and that only two were independent.  The mathematicians
were not asked this question explicitly, but addressed the question
during their discussion of the meaning of the partial derivative.

All of our experts were able to discern the number of degrees of
freedom present in the system, but we were surprised at how long it
took the physicists and engineers to agree upon and express the fact of the
interdependence among the controllable quantities.

\physicists When asked how many independent variables there were in
Excerpt~\ref{exc:phy-dimensions} below, the physicists recognized that
only two variables were independent, because they could independently
control the two forces.
\begin{excerpt}{17:21}\label{exc:phy-dimensions}
\item[INT:]  
  How many independent variables are there?
\item[P2:]  So the way that you defined it, it looks like, so this is the
  question, you know, what exactly is $x$ and what is $y$, right? So if
  this is, if you're saying that this is $y$ [points to $y$ side] and
  this is $x$ [points to $x$ side] and there's only $F_x$ and there's
  only $F_y$, you, you know, I mean...  Uh, otherwise, you know, there's
  also this spring to take into account, so... 
\item[INT:]  So talk to your partner. 
\item[P1:]  Well I would say can could do a way independently on both
  sides. 
\item[P2:]  Right. 
\item[P1:]  So $F_x$ and $F_y$ are independent. 
\item[P2:]  Right. 
\item[P1:]  Anything else we can change. 
\item[P2:]  Right, my only question is this spring [points to spring],
  right? Because, I mean, there's also a force on this side [points to
  $y$ side], right? So the total $F_x$ is also has a contribution from
  the spring [points to spring], in principle. 
\item[P1:]  Umm...  No, because $F_x$ [points towards $F_x$] is the
  tension of this string. 
\item[P2:]  Right, but... 
\item[P1:]  So no matter how much weight I put here [pulling on
  weights], the tension on the spring is the same.  Now I can change $x$
  [points to $x$ position marker] by changing $F_y$ [points to $y$
  string], um, but then that's because $F$ and $y$ are both functions of
  the variables. 
\end{excerpt}
After some more discussion, they agreed that the derivative must be a
function of the two independent variables $F_x$ and $F_y$.

A bit later in Excerpt~\ref{exc:phy-function-three}, the physicists
pondered how to understand the possibility of directly controlling
$y$, suggesting that $x$ depends on three variables each of which they could
control, but they could not control all three independently.  Based on
their gestures leading up to this and the discussion in
Excerpt~\ref{exc:phy-dimensions} above, we believe they arrived at
this conclusion based on ``physical'' reasoning.
\begin{excerpt}{27:05}\label{exc:phy-function-three}
\item[P1:]  Can we say that $x$ is actually a function of $y$, $F_x$,
  and $F_y$, but then these three are not independent. [writes
  $x(y,F_x,F_y)$ on whiteboard] 
\item[P2:]  Exactly. 
\end{excerpt}

\engineers Towards the end of their interview (without interviewer
prompting), the engineers discussed their concept image of a partial
derivative.  In Excerpt~\ref{exc:eng-concept} below, the senior
engineer E1 decided to ask his two partners how they understand the
concept of a partial derivative.  This led to a discussion of how many
independent variables were present in the system.  At this stage we
believe E1 knew the answers to his questions, and had taken on the
role of interviewer of his younger colleagues, who were less certain
as to how to treat this multidimensional system.  Discussion after the
interview confirmed that E1 had been thinking already about how to
employ the PDM in the context of a class in thermodynamics.
\begin{excerpt}{42:19}\label{exc:eng-concept}
\item[E1:] So, let's take a step back, what's your conceptualization of
  the partial derivative? 
\item[E3:] Well it's a great question, E2.
\item[E2:] I just had to ask E3 about it, uh...  It's, I think of it
  as a, in a multidimensional system watching how one, um, dimension
  changes when the others are fixed. 
\item[E1:] Okay, so...  How does that apply to this then? What dimensions
  do we have? 
\item[E2:] Two...  Well we sort of have four. 
\item[E1:] Yeah you kind of do, kind of don't, right? 
\item[E2:] Yeah, I don't know if we're going to go relative
  or... 
\item[E1:] Well, I mean if you didn't know the statics [points to
  spring system] you could say you have four, but you really don't
  have four because how, how this side [points to $x$ side] behaves
  depends on how that side's fixed [points to $y$ side].  And that's
  what E3 was talking about too, right? 
\item[E2:] Umm. 
\item[E1:] So if we're just doing it empirically, we would just say we
  would say we have the length [points to $x$ string] of this the
  $F_x$ of this, the length of that [points to $y$ string] the $F$,
  the tension of that, right? 
\item[E2:] Mhmm. 
\item[E1:] So then, what would the partial derivative be? 
\item[E2:]  The partial derivative of the length with respect to
  force [E1 points at plot of $x$ and $F_x$], well that's helping us.
\item[E1:] The partial derivative of the length with respect to the
  force, 
  we have, since there's four parameters, right?  We
  have our multivariable space. 
\item[E3:] I mean... 
\item[E2:] So I guess, well, the problem is adding them up
  that... when I think of it as an equation, I see more clearly how
  you can manipulate it to more of like some cover, the situation. [E3
  begins writing equation on board] But when we're just making
  measurements, it seems much more incremental. 
\item[E1:] What if we said, so the $x$ is, um, the dependent variable,
  right? So what if we said, what if we just stated $x$ as a function?
  And we would say $x$ here [points to $x$ string] is a function of
  $F_x$, $F_2$ [i.e. $F_y$], and $L_2$ [points to $y$ string].
  \[ x(F_x, F_2, L_2)\]
\end{excerpt}
The engineers (like the physicists) concluded by discussing the
quantity $x$ as a function of the three other variables, and
recognized that those three variables are not themselves independent.

\mathematicians As we described above, the mathematicians were very
quick (once they knew what the variables meant) to recognize that $x$
depends on $F_x$ and $F_y$ (see
Excerpt~\ref{exc:mat-x-of-two-forces}), indicating that they
recognized two independent degrees of freedom.  Moreover, they found
in Excerpt~\ref{exc:mat-clamping-y} that the derivative $\partial
x/\partial F_x$ was smaller if they fixed $y$ rather than fixing
$F_y$, indicating that fixing the third quantity $y$ had some effect.
They did not, however, count the independent degrees of freedom, nor
did we prompt them to do so.

The mathematicians began discussing how to treat the second side of
the system in Excerpt~\ref{exc:mat-mathy-physical}.  M2
expressed a tension between a ``mathy'' understanding and their
physical experimentation which led them to believe that the partial
derivative depended on what was done on the $y$ side of the PDM.
\begin{excerpt}{32:17}\label{exc:mat-mathy-physical}
\item[M2:] But like, ah, I just keep going back to the, you know, more
  mathy idea [points to an illegible derivative on whiteboard], like
  this idea of a partial derivative where, you know, when I'm taking the
  partial derivative of that with respect to $x$, you know $y$ is
  just...  But then we see this like physical thing where this side does
  matter [points to $y$ side], so then... 
\item[M1:] Well do you want to, I mean, can we...  Should we just focus
  on that question and just ask ourselves, like convince ourselves one
  way or the other? 
\item[M2:]  Whether the $y$ matters. 
\end{excerpt}
After agreeing to explore the dependence on the $y$ side of the
system, in Excerpt~\ref{exc:mat-Fy-matters}, the mathematicians
concluded based on experimentation that $x$ is indeed affected by the
value of the other force $F_y$, confirming that the derivative is a
function of two variables.  We note that their language is at times
confusing, as they often use $x$ and $y$ to refer to the two
independent variables (which we call $F_x$ and $F_y$), and had written
$f(x,y)$ on the board to refer to a generic function of two
dimensions.
\begin{excerpt}{32:58}\label{exc:mat-Fy-matters}
\item[M1:]  Okay, so what would the $y$ mattering look like for us?
\item[M2:] So, I guess, the, if the $y$ mattered, um, the force in $y$
  [$F_y$] would have an impact, would somehow have an impact on our $x$
  position, you know, as we...  Here [references $\partial x/\partial F_x$],
  we're trying, we're changing the force in $x$, so with that
  regard, does this force here [points to $y$ side] still have an impact
  when we're just seeing what the change is with regard to that force.
\end{excerpt}
Thus the mathematicians concluded that the partial derivative was a
function of the two forces.

\section{Conclusions: Concept Images and Concept Definitions}

In this section, we discuss the experts' concept images and definitions
for partial derivatives.  The commonalities in the concept images are as
follows (1) a partial derivative is a function; (2) to find a partial
derivative, something must be held fixed; (3) the notion of slope; (4)
the derivative is a ratio of small changes.  There are, however, some
differences: the mathematicians and the engineers were the only ones to
talk about the derivative as a rate of change and as a slope, and the
engineers and mathematicians were the only groups to consider writing an
explicitly defined function.  We also note that having more concept
images is not necessarily ``better.''  Indeed, given the prompt provided
to the experts, measuring and then determining a ratio of small changes
was sufficient to complete the task.  The experts' use of multiple
approaches and images suggests that their concept images are complex yet
they may not have immediately known which image would provide a solution
to the given prompt.

\physicists
The physicists thought of a partial derivative as a ratio
of small changes and as a function.  However, at no point did they
consider writing an explicit function definition; rather, they
attended to determining how many independent variables were in the
system, which of them to fix, and collecting data.

\engineers
Like the physicists, the engineers thought of the partial derivative as
something that could be approximated experimentally, and spent a lot of
time collecting data so that they could represent the partial derivative
as a ratio of small changes.  They also talked about the derivative as a
slope, and that they could graph their data to determine an
explicitly-defined function for the derivative by making use of the
notion of derivative as slope.  They further noted that their
experimentally-determined function should match some sort of theoretical
equation.  The engineers were the only ones to state a concept
definition, albeit incomplete and bordering on inaccurate.  They defined
a partial derivative as ``how one dimension in a multidimensional system
changes when the other dimensions are fixed.''

\mathematicians
The mathematicians repeatedly expressed their interest in determining an
algebraic expression for $\partial x/\partial F_x$; they seemed to
interpret the task statement as directing them to find an explicit
function definition.  Once they had collected data, they talked about the
derivative as a slope, as a ratio of small changes, and as a rate of
change.  Like the physicists, they recognized the need to fix one
quantity, which for them was connected with the idea of holding a
variable constant when using symbolic differentiation rules for a
multivariable function.

\section{Conclusions}

In this section, we summarize the experts' concept images and
definitions for partial derivatives, propose the need for an extended
theoretical framework, describe differences in the concept of limit
between mathematics/theory and experiment, and discuss the limitations
of our study and the prospects for future work.

\subsection{Concept images of derivative}

All three groups of experts assumed that when taking a partial
derivative with respect to $F_x$, $F_y$ can be assumed to be held
fixed.  This finding was unexpected, particularly with regard to those
experts who work with thermodynamics in their research: both
physicists and one engineer.  This question merits further study,
particularly to probe when and how experts approach problems in which
there is ambiguity in the choice of ``independent'' degrees of
freedom.

\physicists
At first, the physicists thought that they were being asked to find (the
inverse of) the spring constant.  Once they recognized that the spring
constant would be represented by a total derivative rather than a
partial derivative, they spent some time exploring other possible
meanings for the interview prompt.  After they were told how to
interpret the prompt, they moved immediately to describing the partial
derivative as a ratio of small numerical changes, collecting data, and
calculating a number.  They spent some time establishing that their
ratio was accurate enough and attended to determining how many
independent variables were in the system and which of them to fix.
While they acknowledged that the derivative was a function, they did not
try to evaluate this function in any way.  They did not mention slope at
all, nor did they try to express the relationship between position and
tension analytically so that they could take a symbolic derivative.

\engineers
Like the physicists, the engineers thought of the partial derivative as
something that could be approximated experimentally, and spent a lot of
time collecting data so that they could represent the partial derivative
as a ratio of small numerical changes.  They mentioned that the
derivative could be found as a slope and that they could therefore
determine the derivative by graphing their data, although they did not
pursue this approach.  They further noted that their
experimentally-determined function should match some sort of theoretical
equation.  The engineers were the only ones to state a concept
definition, albeit incomplete and bordering on inaccurate.  They defined
a partial derivative as ``how one dimension in a multidimensional system
changes when the other dimensions are fixed.''

\mathematicians
The mathematicians repeatedly expressed their interest in determining an
algebraic expression for $\partial x/\partial F_x$; they seemed to
interpret the task statement as directing them to find an explicit
function definition.  Once they had collected data, they talked about the
derivative as a slope, as a ratio of small changes, and as a rate of
change.  Like the physicists, they recognized the need to fix one
quantity, which for them was connected with the idea of holding a
variable constant when using symbolic differentiation rules for a
multivariable function.

\subsection{Need for an Extended Framework}

In section II B, we gave a brief summary of Zandieh's framework for the
concept image of derivative.  In section III B, we gave our own list of
five different ways to understand and think about the concept of
derivative, which has substantial overlap with Zandieh's framework, but
included a specific numerical category.  Throughout our analysis, our
interdisciplinary team found it useful to use elements of both of these
descriptions.  In particular, we found ourselves frequently exploiting
Zandieh's choice of process--object pairs.

However, we had deliberately made the choice to embed our interviews in
an task in which the interviewees had easy access to numerical data, but
not to an analytic expression for the relationship between the physical
quantities.  This is, after all, the environment in which most
experimentalists typically find themselves.  For this reason, our
analysis focused on a numerical representation within the concept image
of derivative.

We were surprised that Zandieh's framework did not include a numerical
representation, even though her introduction made it clear that she was
aware of this possibility.  A careful reading of her paper shows several
examples of students using numbers and these discussions are included in
her analysis.  This suggests that she considers numerical evaluation as
a later step in each of her representations rather than as a
representation in its own right.  In our own list of five categories, we
are now finding the possible need to distinguish between \textit{two}
numerical representations: \textit{exact} numerical work (such as would
happen when plugging particular values for the independent variable into
an explicit analytical model for a physical process) and
\textit{experimental} numerical work (such as data from an experiment,
including its experimental uncertainties).  We are currently working on
a paper which further elucidates these issues.

It will also be necessary to extend any resulting theoretical framework
for the concept image of derivative to elements of the concept of
partial derivative that are distinct from derivative.  In particular, we
see the need to add a layer for what is held fixed when finding a
particular partial derivative.

\subsection{Limits and the Real World}

There were striking differences between the mathematicians and the
physicists and engineers.  The physicists and engineers were happy and
relatively quick to find the derivative at a point as the numerical
ratio of small changes.  In this process, they were so casual with
approximations that they were willing to give this ratio as their
\textit{final answer} to the interview prompt asking for a partial
derivative without qualifying their answer with the word approximation.
They did signal that they understood that their answer was an
approximation by checking that the ratio they calculated was ``good
enough" by either decreasing (physicists) or increasing (engineers) the
change in the independent variable.  They also demonstrated an
understanding of the function layer but couldn't be bothered to find the
value of the derivative at more than one point.  On the other hand, the
mathematicians were firmly embedded in the function layer, pursuing the
desire to find an analytic expression until they were explicitly asked
by the interviewer ``What else could you do?"  Subsequently, when they
were discussing finding a slope from their numerical data, they
continued to doubt the reasonableness of the resulting approximation
(see Excerpt~\ref{exc:mat-small-enough}).

We note that, in the idealized world of pure mathematics (and
theoretical physics), the need to approximate rarely arises, whereas,
for experimentalists, a ratio of small numerical changes is often the
best answer that they have for a derivative, particularly in the absence
of a theoretical model for the process they are studying.  So the
shorthand of calling this ratio ``the derivative" rather than the more
cumbersome ``an approximation to the derivative" makes cultural sense.
Experimentalists can always hope to design a better measuring apparatus
or at least imagine a \textit{gedanken} experiment to improve their
approximation.

In the extreme case, physicists, in particular, are likely aware that
continuity itself and therefore the ability to take formal limits in
derivatives, are properties of the continuous models that they use for
the physical world and not properties of the real world itself on atomic
scales and below.  The real world imposes a fundamental limitation on the
concept of limit.

\subsection{Limitations of this Work}

Obviously, we cannot use a single interview with two or three content
experts to make conclusions about every expert in a given field, nor can
we come to definite conclusions about what these experts may or may not
have been able to do in other settings or contexts.  We plan to
interview different types of mathematicians (especially computational
and applied mathematicians who may have more experience with
approximation), different types of physicists (including
experimentalists and non-computational theorists), and different types
of engineers.  We would also like to interview experts in such fields as
economics or oceanography, whose mathematical cultures may differ
substantially from those already considered (and from each other).  In
particular, we hypothesize that thermodynamics experts from a variety of
fields will behave more similarly on this task than non-thermodynamics
experts within a single field.

\subsection{Other Future Directions}

It would be interesting to ask interviewees how they would report their data
for the partial derivative (assuming they were to take lots of data) if
they were to publish this result in a paper.  We are interested to see
(a)~if they address both dimensions of Zandieh's representation and
(b)~how they would represent the result: tables, single or multiple
one-dimensional plots, three-dimensional surfaces, contours, etc.

We would also be interested to explicitly prompt experts to give us a
concept definition, to see how their concept image (as determined from
their approach to this novel task) relates to their concept definition.

Every group spent some time exploring the physical system of the PDM.
In future, it would be interesting to explore this aspect of the
interviews and to think about the pedagogical and research implications
of this type of play as an aspect of expert reasoning.  In particular,
the mathematicians spent more time than other groups exploring the
covariation of the physical quantities by pulling on various strings and
making expert-like observations of how the system responded.

Finally, we are interested in pursuing the pedagogical consequences of
this study for classroom learning trajectories.  For example,
physics students have traditionally been exposed to the discreteness of
data and the necessity of addressing experimental uncertainty in their
lower division laboratories.  But this exposure is disappearing with the
near ubiquitous use of motion sensors and computer interfacing that
blur the distinction between discrete data and continuous plots.  What
other physics experiences (research, advanced laboratory courses, etc.)
will be needed to reinforce these aspects of the concept image of
derivative?

\acknowledgments{
  We especially wish to acknowledge the helpful conversations we had
  regarding this paper with Joseph F. Wagner and Emily Smith.
  The funding for this project was provided, in part, by the National
  Science Foundation under Grant Nos.  DUE 0618877, DUE 0837829,
  DUE 1023120, and DUE 1323800.
}

\bibliography{prstper}

\end{document}